\documentclass[11pt,a4paper]{article}
\usepackage{amsmath}

\begin{document}

\title{{\textbf{\large Mermin's \textit{\textbf{n}}-particle Bell inequality and operators' noncommutativity}}}
\vspace{1cm}
\author{Jos\'{e} L. Cereceda\thanks{Electronic mail: jl.cereceda@teleline.es} \\
\textit{C/Alto del Le\'{o}n 8, 4A, 28038 Madrid, Spain}}

\date{Published on 13 August 2001}

\maketitle

\begin{abstract}
The relationship between the noncommutativity of operators and the violation of the Bell inequality is exhibited in the light of the $n$-particle Bell-type inequality discovered by Mermin [Phys. Rev. Lett. 65 (1990) 1838]. It is shown, in particular, that the maximal amount of violation of Mermin's inequality predicted by quantum mechanics decreases exponentially by a factor of $2^{-m/2}$ whenever any $m$ among the $n$ single-particle commutators happen to vanish.

\vspace{.5cm}
\end{abstract}

\newpage

The quantum-mechanical violation of the Clauser-Horne-Shimony-Holt (CHSH) form [1] of Bell's inequality [2] is a direct consequence of the fact that quantum operators obey a noncommutative algebra. For two spin-$\tfrac{1}{2}$ particles, the CHSH inequality can be written as $|\langle \hat{\mathcal{B}}_{\text{CHSH}} \rangle | \leq 2$, where $\langle \hat{\mathcal{B}}_{\text{CHSH}} \rangle$ denotes the expectation value of the Bell operator [3]
\begin{equation}
\hat{\mathcal{B}}_{\text{CHSH}} =\sigma({\hat{\textbf {n}}}_1)\sigma({\hat{\textbf {n}}}_2)
+ \sigma({\hat{\textbf {n}}}_1)\sigma({\hat{\textbf {n}}}_{2}^{\prime})
+ \sigma({\hat{\textbf {n}}}_{1}^{\prime})\sigma({\hat{\textbf {n}}}_2)
- \sigma({\hat{\textbf {n}}}_{1}^{\prime})\sigma({\hat{\textbf {n}}}_{2}^{\prime})\, .
\end{equation}
In Eq.\ (1), $\sigma({\hat{\textbf {n}}}_j)$ ($\sigma({\hat{\textbf {n}}}_{j}^{\prime})$) denotes the spin operator for particle $j$ along the direction ${\hat{\textbf {n}}}_j$ (${\hat{\textbf {n}}}_{j}^{\prime}$). From the fact that $\sigma^2({\hat{\textbf {n}}}_1)= \sigma^2({\hat{\textbf {n}}}_{1}^{\prime})= \sigma^2({\hat{\textbf {n}}}_2)= \sigma^2({\hat{\textbf {n}}}_{2}^{\prime})= \hat{I}$, it follows immediately that [4]
\begin{equation}
\hat{\mathcal{B}}_{\text{CHSH}}^2 = 4\hat{I} - [\sigma({\hat{\textbf {n}}}_1),\sigma({\hat{\textbf {n}}}_{1}^{\prime})] [\sigma({\hat{\textbf {n}}}_2),\sigma({\hat{\textbf {n}}}_{2}^{\prime})] \, .
\end{equation}
From Eq.\ (2), we can see that the CHSH inequality may be violated by the quantum-mechanical predictions provided that $[\sigma({\hat{\textbf {n}}}_1),\sigma({\hat{\textbf {n}}}_{1}^{\prime})] \neq 0$ and $[\sigma({\hat{\textbf {n}}}_2),\sigma({\hat{\textbf {n}}}_{2}^{\prime})] \neq 0$. Whenever either commutator $[\sigma({\hat{\textbf {n}}}_1),\sigma({\hat{\textbf {n}}}_{1}^{\prime})]$ or $[\sigma({\hat{\textbf {n}}}_2),\sigma({\hat{\textbf {n}}}_{2}^{\prime})]$ vanishes then $\hat{\mathcal{B}}_{\text{CHSH}} = \pm 2\hat{I}$, and so $|\langle \hat{\mathcal{B}}_{\text{CHSH}} \rangle| = 2$ for any joint state describing the spin of the particles. On the other hand, for three spin-$\tfrac{1}{2}$ particles the appropriate Bell inequality is of the form $|\langle \hat{\mathcal{B}}_{\text{H}} \rangle | \leq 2$ (see Eq.\ (14) of [5]), where now the representative Bell operator can be written as
\begin{equation}
\hat{\mathcal{B}}_{\text{H}} = \sigma({\hat{\textbf {n}}}_{1}^{\prime})\sigma({\hat{\textbf {n}}}_{2})
\sigma({\hat{\textbf {n}}}_{3}) + \sigma({\hat{\textbf {n}}}_{1})\sigma({\hat{\textbf {n}}}_{2}^{\prime})
\sigma({\hat{\textbf {n}}}_{3})  
+ \sigma({\hat{\textbf {n}}}_{1})\sigma({\hat{\textbf {n}}}_{2})
\sigma({\hat{\textbf {n}}}_{3}^{\prime}) - \sigma({\hat{\textbf {n}}}_{1}^{\prime})\sigma({\hat{\textbf {n}}}_{2}^{\prime})
\sigma({\hat{\textbf {n}}}_{3}^{\prime})  \, .
\end{equation}
It is simple algebra to verify that [6]
\begin{multline}
\qquad \hat{\mathcal{B}}_{\text{H}}^2 = 4\hat{I} - [\sigma({\hat{\textbf {n}}}_1),
\sigma({\hat{\textbf {n}}}_{1}^{\prime})] [\sigma({\hat{\textbf {n}}}_2),\sigma({\hat{\textbf {n}}}_{2}^{\prime})]  \\
- [\sigma({\hat{\textbf {n}}}_2),\sigma({\hat{\textbf {n}}}_{2}^{\prime})] 
[\sigma({\hat{\textbf {n}}}_3),\sigma({\hat{\textbf {n}}}_{3}^{\prime})] 
- [\sigma({\hat{\textbf {n}}}_1),\sigma({\hat{\textbf {n}}}_{1}^{\prime})] 
[\sigma({\hat{\textbf {n}}}_3),\sigma({\hat{\textbf {n}}}_{3}^{\prime})] \, . \qquad
\end{multline}
From Eq.\ (4), we can see that in order for the inequality $|\langle \hat{\mathcal{B}}_{\text{H}} \rangle | \leq 2$ to be violated by the quantum predictions it is necessary that at least two of the commutators involved be nonzero. When one of the commutators, say $[\sigma({\hat{\textbf {n}}}_3),\sigma({\hat{\textbf {n}}}_{3}^{\prime})]$, vanishes then the expression in Eq.\ (4) reduces to $\hat{\mathcal{B}}_{\text{H}}^2 = 4\hat{I}  - [\sigma({\hat{\textbf {n}}}_1),\sigma({\hat{\textbf {n}}}_{1}^{\prime})] [\sigma({\hat{\textbf {n}}}_2),\sigma({\hat{\textbf {n}}}_{2}^{\prime})]$ which corresponds to the square of the two-particle Bell operator $\hat{\mathcal{B}}_{\text{CHSH}}$, Eq.\ (2). This means that whenever the commutator $[\sigma({\hat{\textbf {n}}}_j),\sigma({\hat{\textbf {n}}}_{j}^{\prime})]$ associated with any one of the particles $j$, $j=1,2,3$, is equal to zero then this particle plays no role in the violation of the corresponding three-particle Bell inequality, so that this Bell inequality of order 3 reduces to a Bell inequality of order 2, where by order of a Bell inequality we want to mean the number of ``effective'' particles involved in the associated Bell operator. The maximum eigenvalue of the operator $\hat{\mathcal{B}}_{\text{CHSH}}$ is (in terms of the absolute value) $2\sqrt{2}$, whereas the maximum eigenvalue of $\hat{\mathcal{B}}_{\text{H}}$ is 4. Consequently, the maximal quantum-mechanical violation of the inequality $|\langle \hat{\mathcal{B}}_{\text{H}} \rangle | \leq 2$ decreases by a factor of $2\sqrt{2}/4 =1/\sqrt{2}$ when any one of the three commutators vanishes. (Of course, for two vanishing commutators we have $\hat{\mathcal{B}}_{\text{H}} = \pm 2\hat{I}$, and then the inequality at issue is saturated by the quantum-mechanical predictions.)

In this Letter, I would like to extend these results further to the general case in which $n$ spin-$\tfrac{1}{2}$ particles are considered. We will show that, under certain specific conditions which will be precisely stated below, a Bell inequality of order $n$ reduces to a Bell inequality of order $n-m$ whenever any $m$ of the $n$ single-particle commutators $[\sigma({\hat{\textbf {n}}}_j),\sigma({\hat{\textbf {n}}}_{j}^{\prime})]$ vanish. This is done in a way that explicitly shows the recognised relation [7,8] between the presence of commutators in the Bell operator and the peculiar nonlocal properties exhibited by quantum mechanics. For this purpose we consider the Bell-type inequality derived by Mermin in 1990 [9] for a system of $n$ spin-$\tfrac{1}{2}$ particles ($n \geq 3$) in the Greenberger-Horne-Zeilinger state [10,11]
\begin{equation}
|\Phi \rangle =(1/\sqrt{2}) \left(|\uparrow \uparrow \cdots \uparrow \rangle
+ i |\downarrow \downarrow \cdots \downarrow \rangle \right),
\end{equation}
where $\uparrow$ or $\downarrow$ in the $j$th position corresponds to the component of the spin of the $j$th particle along its own $z$ axis. Mermin's Bell inequality can be summarised as [9]
\begin{align}
|\langle \hat{\mathcal{B}}_{\text{M}} \rangle | & \leq 2^{n/2},
\quad \text{$n$ even},  \tag{6a}  \\
|\langle \hat{\mathcal{B}}_{\text{M}} \rangle | & \leq 2^{(n-1)/2},
\quad \text{$n$ odd},  \tag{6b}
\setcounter{equation}{6}
\end{align}
where $\langle\hat{\mathcal{B}}_{\text{M}}\rangle$ denotes the expectation value of the (Hermitian) Bell operator
\begin{equation}
\hat{\mathcal{B}}_{\text{M}} = \frac{1}{2i} \left(
\prod_{j=1}^{n} (\sigma_x^j + i\sigma_y^j )
- \prod_{j=1}^{n} (\sigma_x^j - i\sigma_y^j )  \right).
\end{equation}
Mermin showed that the prediction that quantum mechanics makes for state (5) violates the inequalities in Eqs.\ (6) by an exponentially large factor of $2^{(n-2)/2}$ for $n$ even or $2^{(n-1)/2}$ for $n$ odd. This was the first spectacular demonstration of the fact that there is no limit to the amount by which the quantum-mechanical correlations can exceed the limits imposed by a Bell inequality.\footnote{%
A noteworthy alternative derivation of the locally realistic bounds in Eqs.\ (6), along with the  bound prescribed by quantum mechanics, can be found in Ref.\ [12] (see, in particular, the equations (3.15), (3.17), and (3.18) of [12]).} In order to achieve a greater generality, we will consider the most general form of the Bell operator in Eq.\ (7), namely,
\begin{equation}
\hat{\mathcal{B}}_{\text{M}} = \frac{1}{2i} \left(
\prod_{j=1}^{n} (\sigma({\hat{\textbf {n}}}_j) + i\sigma({\hat{\textbf {n}}}_{j}^{\prime}))
- \prod_{j=1}^{n} (\sigma({\hat{\textbf {n}}}_j) - i\sigma({\hat{\textbf {n}}}_{j}^{\prime}))
 \right),
\end{equation}
where ${\hat{\textbf {n}}}_j$ and ${\hat{\textbf {n}}}_{j}^{\prime}$ are arbitrary directions. Of course, the Mermin's Bell inequalities (6) still hold for the general Bell operator (8). The treatment that follows does not rely on any particular state like that of Eq.\ (5). Rather, it is based on the general properties exhibited by the square of the Bell operator (8) when expressed in terms of the commutators $[\sigma({\hat{\textbf {n}}}_j),\sigma({\hat{\textbf {n}}}_{j}^{\prime})]$. In order to abbreviate the notation, we shall henceforth drop the $\sigma$'s of the commutators so that they will be written simply as $[{\hat{\textbf {n}}}_j, {\hat{\textbf {n}}}_{j}^{\prime}]$. Now we must distinguish between the cases of $n$ odd and $n$ even. For an odd number of particles ($n \geq 3$) the square of operator (8) is given by
\begin{multline}
\hat{\mathcal{B}}_{\text{M}}^{2}(\text{$n$ odd})=2^{n-1} \hat{I}
- 2^{n-3}\, \overbrace{\sum_{m_1 < m_2} [{\hat{\textbf {n}}}_{m_1},{\hat{\textbf {n}}}_{m_1}^{\prime}]
[{\hat{\textbf {n}}}_{m_2},{\hat{\textbf {n}}}_{m_2}^{\prime}]}^{\left( \begin{smallmatrix} n \\ 2 \end{smallmatrix} \right) \text{ terms}}  \\
+2^{n-5}\, \overbrace{\sum_{m_1 < m_2 < m_3 < m_4} [{\hat{\textbf {n}}}_{m_1},{\hat{\textbf {n}}}_{m_1}^{\prime}]
[{\hat{\textbf {n}}}_{m_2},{\hat{\textbf {n}}}_{m_2}^{\prime}] [{\hat{\textbf {n}}}_{m_3},{\hat{\textbf {n}}}_{m_3}^{\prime}]
[{\hat{\textbf {n}}}_{m_4},{\hat{\textbf {n}}}_{m_4}^{\prime}]}^{\left(\begin{smallmatrix} n \\ 4 \end{smallmatrix}
\right) \text{ terms}} \\
+\ldots + (-1)^k \, 2^{n-2k-1}\, \overbrace{\sum_{m_1 < m_2 < \ldots < m_{2k}} [{\hat{\textbf {n}}}_{m_1},{\hat{\textbf {n}}}_{m_1}^{\prime}] [{\hat{\textbf {n}}}_{m_2},{\hat{\textbf {n}}}_{m_2}^{\prime}] \cdots [{\hat{\textbf {n}}}_{m_{2k}},{\hat{\textbf {n}}}_{m_{2k}}^{\prime}]}^{\left(\begin{smallmatrix}
n \\ 2k \end{smallmatrix} \right) \text{ terms}}  \\
+\ldots + (-1)^{(n-1)/2}\, \overbrace{\sum_{m_1 < m_2 < \ldots < m_{n-1}} [{\hat{\textbf {n}}}_{m_1},{\hat{\textbf {n}}}_{m_1}^{\prime}] [{\hat{\textbf {n}}}_{m_2},{\hat{\textbf {n}}}_{m_2}^{\prime}] \cdots [{\hat{\textbf {n}}}_{m_{n-1}},{\hat{\textbf {n}}}_{m_{n-1}}^{\prime}]}^{\left(\begin{smallmatrix}
n \\ n-1 \end{smallmatrix} \right) \text{ terms}}\, ,
\end{multline}
where \{$m_1,m_2,\ldots,m_{n-1}\}$ is a set of $n-1$ indices each of which running from 1 to $n$. Analogously, for $n$ even ($n \geq 4$) we have the slightly more complicated expression
\begin{multline}
\hat{\mathcal{B}}_{\text{M}}^{2}(\text{$n$ even})=2^{n-1} \hat{I}
- 2^{n-3}\, \overbrace{\sum_{m_1 < m_2} [{\hat{\textbf {n}}}_{m_1},{\hat{\textbf {n}}}_{m_1}^{\prime}]
[{\hat{\textbf {n}}}_{m_2},{\hat{\textbf {n}}}_{m_2}^{\prime}]}^{\left(\begin{smallmatrix} n \\ 2 \end{smallmatrix}
\right) \text{ terms}}  \\
+2^{n-5}\, \overbrace{\sum_{m_1 < m_2 < m_3 < m_4} [{\hat{\textbf {n}}}_{m_1},{\hat{\textbf {n}}}_{m_1}^{\prime}]
[{\hat{\textbf {n}}}_{m_2},{\hat{\textbf {n}}}_{m_2}^{\prime}] [{\hat{\textbf {n}}}_{m_3},{\hat{\textbf {n}}}_{m_3}^{\prime}]
[{\hat{\textbf {n}}}_{m_4},{\hat{\textbf {n}}}_{m_4}^{\prime}]}^{\left(\begin{smallmatrix} n \\ 4 \end{smallmatrix}
\right) \text{ terms}} \\
+\ldots + (-1)^k \, 2^{n-2k-1}\, \overbrace{\sum_{m_1 < m_2 < \ldots < m_{2k}} [{\hat{\textbf {n}}}_{m_1},{\hat{\textbf {n}}}_{m_1}^{\prime}] [{\hat{\textbf {n}}}_{m_2},{\hat{\textbf {n}}}_{m_2}^{\prime}] \cdots [{\hat{\textbf {n}}}_{m_{2k}},{\hat{\textbf {n}}}_{m_{2k}}^{\prime}]}^{\left(\begin{smallmatrix}
n \\ 2k \end{smallmatrix} \right) \text{ terms}}  \\
+\ldots + (-1)^{(n-2)/2}\, 2\, \overbrace{\sum_{m_1 < m_2 < \ldots < m_{n-2}} [{\hat{\textbf {n}}}_{m_1},{\hat{\textbf {n}}}_{m_1}^{\prime}] [{\hat{\textbf {n}}}_{m_2},{\hat{\textbf {n}}}_{m_2}^{\prime}] \cdots [{\hat{\textbf {n}}}_{m_{n-2}},{\hat{\textbf {n}}}_{m_{n-2}}^{\prime}]}^{\left(\begin{smallmatrix}
n \\ n-2 \end{smallmatrix} \right) \text{ terms}}  \\
+ (-1)^{n/2}\,\frac{1}{2} \left( [{\hat{\textbf {n}}}_1,{\hat{\textbf {n}}}_1^{\prime}] [{\hat{\textbf {n}}}_2,{\hat{\textbf {n}}}_2^{\prime}] \cdots [{\hat{\textbf {n}}}_n,{\hat{\textbf {n}}}_n^{\prime}] - \{ {\hat{\textbf {n}}}_1,{\hat{\textbf {n}}}_1^{\prime}\} \{ {\hat{\textbf {n}}}_2,{\hat{\textbf {n}}}_2^{\prime}\} \cdots \{ {\hat{\textbf {n}}}_n,{\hat{\textbf {n}}}_n^{\prime}\} \right)  ,
\end{multline}
where $\{m_1,m_2,\ldots,m_{n-2}\}$ is a set of $n-2$ indices each of which running from 1 to $n$, and $\{ {\hat{\textbf {n}}}_j,{\hat{\textbf {n}}}_j^{\prime}\}$ is the anticommutator of the operators $\sigma({\hat{\textbf {n}}}_j)$ and $\sigma({\hat{\textbf {n}}}_j^{\prime})$.

Let us first examine what happens when any \textit{two\/} of the commutators in Eq.\ (9) are equal to zero. In the first place, it is clear that all last $\binom{n}{n-1}$ terms in Eq.\ (9) vanish when any two commutators are equal to zero, since each of these $n$ terms is a product of $n-1$ distinct commutators. Furthermore, it can be seen that, for each group of $\binom{n}{2k}$ terms in Eq.\ (9) (with each of the terms in a group being a product of $2k$ commutators, $2k\leq n-3$), there is a total of $\binom{n-2}{2k}$ terms which remain ``untouched'' when any two of the commutators are made to vanish, the remaining $\binom{n}{2k}-\binom{n-2}{2k}$ terms in the group being equal to zero due to the vanishing of the two commutators. Therefore, whenever any two commutators are equal to zero, expression (9) reduces to
\begin{multline}
\hat{\mathcal{B}}_{\text{M}}^{2}(\text{$n$ odd})\to 2^2 \Bigl( 2^{n-3} \hat{I}
- 2^{n-5}\, \overbrace{\sum_{m_1 < m_2} [{\hat{\textbf {n}}}_{m_1},{\hat{\textbf {n}}}_{m_1}^{\prime}]
[{\hat{\textbf {n}}}_{m_2},{\hat{\textbf {n}}}_{m_2}^{\prime}]}^{\left( \begin{smallmatrix} n-2 \\ 2 \end{smallmatrix}
\right) \text{ terms}}  \\
+2^{n-7}\, \overbrace{\sum_{m_1 < m_2 < m_3 < m_4} [{\hat{\textbf {n}}}_{m_1},{\hat{\textbf {n}}}_{m_1}^{\prime}]
[{\hat{\textbf {n}}}_{m_2},{\hat{\textbf {n}}}_{m_2}^{\prime}] [{\hat{\textbf {n}}}_{m_3},{\hat{\textbf {n}}}_{m_3}^{\prime}]
[{\hat{\textbf {n}}}_{m_4},{\hat{\textbf {n}}}_{m_4}^{\prime}]}^{\left(\begin{smallmatrix} n-2 \\ 4 \end{smallmatrix}
\right) \text{ terms}} \\
+\ldots + (-1)^k \, 2^{n-2k-3}\, \overbrace{\sum_{m_1 < m_2 < \ldots < m_{2k}} [{\hat{\textbf {n}}}_{m_1},{\hat{\textbf {n}}}_{m_1}^{\prime}] [{\hat{\textbf {n}}}_{m_2},{\hat{\textbf {n}}}_{m_2}^{\prime}] \cdots [{\hat{\textbf {n}}}_{m_{2k}},{\hat{\textbf {n}}}_{m_{2k}}^{\prime}]}^{\left(\begin{smallmatrix}
n-2 \\ 2k \end{smallmatrix} \right) \text{ terms}}  \\
+\ldots + (-1)^{(n-3)/2}\, \overbrace{\sum_{m_1 < m_2 < \ldots < m_{n-3}} [{\hat{\textbf {n}}}_{m_1},{\hat{\textbf {n}}}_{m_1}^{\prime}] [{\hat{\textbf {n}}}_{m_2},{\hat{\textbf {n}}}_{m_2}^{\prime}] \cdots [{\hat{\textbf {n}}}_{m_{n-3}},{\hat{\textbf {n}}}_{m_{n-3}}^{\prime}]}^{\left(\begin{smallmatrix}
n-2 \\ n-3 \end{smallmatrix} \right) \text{ terms}} \, \Bigr) .
\end{multline}
We may enumerate the $n-2$ particles corresponding to the $n-2$ nonvanishing commutators, so that each of the $n-3$ indices $\{m_1,m_2,\ldots,m_{n-3}\}$ in Eq.\ (11) runs from 1 to $n-2$. We introduce the notation $\hat{\mathcal{B}}_{\text{M}}^2 (n|m)$ to denote the squared $n$-particle Bell operator that obtains when $m$ single-particle commutators vanish. Then we can put the whole expression (11) in a compact notation as $\hat{\mathcal{B}}_{\text{M}}^2 (\text{$n$ odd}|2)= 2^2 \hat{\mathcal{B}}_{\text{M}}^2 (n-2)$. Suppose now that any two commutators appearing in Eq. (11) happen in turn to vanish. Then it is clear that an analysis similar to the one we have just performed for Eq.\ (9) would enable us to deduce that $\hat{\mathcal{B}}_{\text{M}}^2 (\text{$n$ odd}|4)= 2^4 \hat{\mathcal{B}}_{\text{M}}^2 (n-4)$. Iterating this procedure successively would lead us to finally conclude that
\begin{equation}
\hat{\mathcal{B}}_{\text{M}}^2 (\text{$n$ odd}|2k)= 2^{2k} \hat{\mathcal{B}}_{\text{M}}^2 
(n-2k), \quad 2k=0,2,4,\ldots,n-3 .   
\end{equation}

Likewise, we may determine the expression which results when two of the commutators in Eq.\ (10) are equal to zero. To do this, we need the auxiliary result according to which, for a spin-$\tfrac{1}{2}$ particle, if $[{\hat{\textbf {n}}}_j, {\hat{\textbf {n}}}_{j}^{\prime}]=0$ then necessarily the anticommutator $\{ {\hat{\textbf {n}}}_j, {\hat{\textbf {n}}}_{j}^{\prime} \}$ is either $2{\hat{I}}_j$ or $-2{\hat{I}}_j$, with ${\hat{I}}_j$ being the identity operator acting on the Hilbert space pertaining to particle $j$. Without any loss of generality, we shall take the axes ${\hat{\textbf {n}}}_i$, ${\hat{\textbf {n}}}_{i}^{\prime}$, ${\hat{\textbf {n}}}_j$, and ${\hat{\textbf {n}}}_{j}^{\prime}$, $i\neq j$, such that whenever we have $[{\hat{\textbf {n}}}_i, {\hat{\textbf {n}}}_{i}^{\prime}]= [{\hat{\textbf {n}}}_j, {\hat{\textbf {n}}}_{j}^{\prime}]=0$ then $\{ {\hat{\textbf {n}}}_i, {\hat{\textbf {n}}}_{i}^{\prime} \} \{ {\hat{\textbf {n}}}_j, {\hat{\textbf {n}}}_{j}^{\prime} \} = -4{\hat{I}}_i {\hat{I}}_j$. Then it can be shown that, for any two vanishing commutators, the expression in Eq.\ (10) reduces to
\begin{multline}
\hat{\mathcal{B}}_{\text{M}}^{2}(\text{$n$ even})\to 2^2 \Bigl( 2^{n-3} \hat{I}
- 2^{n-5}\, \overbrace{\sum_{m_1 < m_2} [{\hat{\textbf {n}}}_{m_1},{\hat{\textbf {n}}}_{m_1}^{\prime}]
[{\hat{\textbf {n}}}_{m_2},{\hat{\textbf {n}}}_{m_2}^{\prime}]}^{\left(\begin{smallmatrix} n-2 \\ 2 \end{smallmatrix}
\right) \text{ terms}}  \\
+2^{n-7}\, \overbrace{\sum_{m_1 < m_2 < m_3 < m_4} [{\hat{\textbf {n}}}_{m_1},{\hat{\textbf {n}}}_{m_1}^{\prime}]
[{\hat{\textbf {n}}}_{m_2},{\hat{\textbf {n}}}_{m_2}^{\prime}] [{\hat{\textbf {n}}}_{m_3},{\hat{\textbf {n}}}_{m_3}^{\prime}]
[{\hat{\textbf {n}}}_{m_4},{\hat{\textbf {n}}}_{m_4}^{\prime}]}^{\left(\begin{smallmatrix} n-2 \\ 4 \end{smallmatrix}
\right) \text{ terms}} \\
+\ldots + (-1)^k \, 2^{n-2k-3}\, \overbrace{\sum_{m_1 < m_2 < \ldots < m_{2k}} [{\hat{\textbf {n}}}_{m_1},{\hat{\textbf {n}}}_{m_1}^{\prime}] [{\hat{\textbf {n}}}_{m_2},{\hat{\textbf {n}}}_{m_2}^{\prime}] \cdots [{\hat{\textbf {n}}}_{m_{2k}},{\hat{\textbf {n}}}_{m_{2k}}^{\prime}]}^{\left(\begin{smallmatrix}
n-2 \\ 2k \end{smallmatrix} \right) \text{ terms}}  \\
+\ldots + (-1)^{(n-4)/2}\, 2\, \overbrace{\sum_{m_1 < m_2 < \ldots < m_{n-4}} [{\hat{\textbf {n}}}_{m_1},{\hat{\textbf {n}}}_{m_1}^{\prime}] [{\hat{\textbf {n}}}_{m_2},{\hat{\textbf {n}}}_{m_2}^{\prime}] \cdots [{\hat{\textbf {n}}}_{m_{n-4}},{\hat{\textbf {n}}}_{m_{n-4}}^{\prime}]}^{\left(\begin{smallmatrix}
n-2 \\ n-4 \end{smallmatrix} \right) \text{ terms}}  \\
+ (-1)^{(n-2)/2}\,\frac{1}{2} \left( [{\hat{\textbf {n}}}_1,{\hat{\textbf {n}}}_1^{\prime}] [{\hat{\textbf {n}}}_2,{\hat{\textbf {n}}}_2^{\prime}] \cdots [{\hat{\textbf {n}}}_{n-2},{\hat{\textbf {n}}}_{n-2}^{\prime}] - \{ {\hat{\textbf {n}}}_1,{\hat{\textbf {n}}}_1^{\prime}\} \{ {\hat{\textbf {n}}}_2,{\hat{\textbf {n}}}_2^{\prime}\} \cdots \{ {\hat{\textbf {n}}}_{n-2},{\hat{\textbf {n}}}_{n-2}^{\prime}\} \right) \Bigr)  ,
\end{multline}
where the particles corresponding to the $n-2$ nonvanishing commutators have been enumerated so that each of the $n-4$ indices $\{m_1,m_2,\ldots,m_{n-4}\}$ in Eq.\ (13) runs from 1 to $n-2$. Eq.\ (13) can be expressed as $\hat{\mathcal{B}}_{\text{M}}^2 (\text{$n$ even}|2)= 2^2 \hat{\mathcal{B}}_{\text{M}}^2 (n-2)$. If an even number $2k$ of commutators are equal to zero then, by applying successively the reasoning leading to Eq.\ (13), we could equally conclude that (cf.\ Eq.\ (12))
\begin{equation}
\hat{\mathcal{B}}_{\text{M}}^2 (\text{$n$ even}|2k)= 2^{2k} \hat{\mathcal{B}}_{\text{M}}^2 
(n-2k), \quad 2k=0,2,4,\ldots,n-4 .   
\end{equation}

A natural question that arises is whether the squared Bell operator $\hat{\mathcal{B}}_{\text{M}}^2 (n)$ does ``collapse'' into $\hat{\mathcal{B}}_{\text{M}}^2 (n-1)$ when one of the commutators is equal to zero. A glance at expressions (9) and (10) tells us that $\hat{\mathcal{B}}_{\text{M}}^2 (n)$ cannot, in general, reduce to $\hat{\mathcal{B}}_{\text{M}}^2 (n-1)$ because of the presence of the product of anticommutators  in the last term of Eq.\ (10). If this product were equal to zero then there  would be a ``continuous'' transition between $\hat{\mathcal{B}}_{\text{M}}^2 (n)$ and $\hat{\mathcal{B}}_{\text{M}}^2 (n-1)$ when one commutator vanishes or, more generally, between 
$\hat{\mathcal{B}}_{\text{M}}^2 (n)$ and $\hat{\mathcal{B}}_{\text{M}}^2 (n-m)$ when $m$ commutators vanish. Obviously, in order for the product $\{ {\hat{\textbf {n}}}_1,{\hat{\textbf {n}}}_1^{\prime}\} \{ {\hat{\textbf {n}}}_2,{\hat{\textbf {n}}}_2^{\prime}\} \cdots \{ {\hat{\textbf {n}}}_n,{\hat{\textbf {n}}}_n^{\prime}\}$ to be equal to zero, it suffices that any given one of the factors vanishes. So we shall assume that one of the anticommutators, $\{ {\hat{\textbf {n}}}_i,{\hat{\textbf {n}}}_{i}^{\prime}\}$ say, is zero, which amounts to making the directions ${\hat{\textbf {n}}}_i$ and ${\hat{\textbf {n}}}_{i}^{\prime}$ perpendicular between themselves. Regarding the quantum-mechanical violation of the Mermin's Bell inequalities (6), the restriction
$\{ {\hat{\textbf {n}}}_i,{\hat{\textbf {n}}}_{i}^{\prime}\}=0$ does not entail any real limitation. In fact, as we shall see below, in order to achieve the maximum quantum violation of Mermin's inequalities, it is necessary that ${\hat{\textbf {n}}}_j$ be perpendicular to ${\hat{\textbf {n}}}_{j}^{\prime}$ for \textit{all\/} $j=1,2,\ldots,n$. At any event, for the case in which one of the anticommutators vanishes, an analysis similar to that used to derive the Eqs.\ (12) and (14) would enable us to conclude that
\begin{equation}
\hat{\mathcal{B}}_{\text{M}}^2 (\text{$n$ odd}|2k+1)= 2^{2k+1} \hat{\mathcal{B}}_{\text{M}}^2 
(n-2k-1), \quad 2k+1=1,3,\ldots,n-4 ,   
\end{equation}
and
\begin{equation}
\hat{\mathcal{B}}_{\text{M}}^2 (\text{$n$ even}|2k+1)= 2^{2k+1} \hat{\mathcal{B}}_{\text{M}}^2 
(n-2k-1), \quad 2k+1=1,3,\ldots,n-3 .   
\end{equation}
We stress that the condition $\{ {\hat{\textbf {n}}}_i,{\hat{\textbf {n}}}_{i}^{\prime}\}=0$ upon which relations (15) and (16) are based, restricts in no way their ``practical'' validity since, as can be seen from Eqs.\ (9) and (10), it is absolutely necessary that at least two of the commutators be nonzero, if we want that the quantum-mechanical predictions can violate inequalities (6). Hence, as the condition $\{ {\hat{\textbf {n}}}_i,{\hat{\textbf {n}}}_{i}^{\prime}\}=0$ implies that $[{\hat{\textbf {n}}}_i,{\hat{\textbf {n}}}_{i}^{\prime}]\neq 0$, we can always choose one of the nonvanishing commutators to be $[{\hat{\textbf {n}}}_i,{\hat{\textbf {n}}}_{i}^{\prime}]$.

Eqs.\ (12) and (14)-(16) can be unified in the single relation
\begin{equation}
\hat{\mathcal{B}}_{\text{M}}^2 (n|m)= 2^m \hat{\mathcal{B}}_{\text{M}}^2 
(n-m), \quad m=0,1,2,\ldots,n-3 ,   
\end{equation}
which applies for any $n$ ($n\geq 3$).\footnote{%
The case $m=n-2$ is excluded because the Mermin's Bell operator $\hat{\mathcal{B}}_{\text{M}}(2)=\sigma({\hat{\textbf {n}}}_1)\sigma({\hat{\textbf {n}}}_{2}^{\prime}) + \sigma({\hat{\textbf {n}}}_{1}^{\prime})\sigma({\hat{\textbf {n}}}_2)$ does not entail any meaningful Bell-type inequality. The transition from the three-particle Bell operator $\hat{\mathcal{B}}_{\text{H}}$ (which can be identified with $\hat{\mathcal{B}}_{\text{M}}(3)$) to the two-particle Bell operator $\hat{\mathcal{B}}_{\text{CHSH}}$ has been treated separately at the opening part of this Letter.} So, as the operator $\hat{\mathcal{B}}_{\text{M}}^2 (n|m)$ is proportional to \mbox{$\hat{\mathcal{B}}_{\text{M}}^2 (n-m)$}, we have shown that a Bell inequality of order $n$ reduces to a Bell inequality of order \mbox{$n-m$} whenever any $m$ of the $n$ single-particle commutators $[{\hat{\textbf {n}}}_j,{\hat{\textbf {n}}}_{j}^{\prime}]$ are equal to zero. The maximum eigenvalue of the Bell operator $\hat{\mathcal{B}}_{\text{M}}(n)$ is $2^{n-1}$. From Eq.\ (17), we can see that the maximum eigenvalue of the operator $\hat{\mathcal{B}}_{\text{M}}(n|m)$ is $2^{m/2}$ times the maximum eigenvalue of $\hat{\mathcal{B}}_{\text{M}}(n-m)$, namely $2^{-m/2} 2^{n-1}$. It thus follows that for the case in which $m$ single-particle commutators vanish, the maximum amount of violation predicted by quantum mechanics of the $n$-particle Bell-type inequalities (6) diminishes by a factor of $2^{-m/2}$ with respect to the maximal violation obtained when \textit{all $n$ anticommutators\/} vanish. In particular, for one vanishing commutator, the decrease factor is $1/\sqrt{2}$.

We conclude by noting an important feature of the general expressions in Eqs.\ (9) and (10), namely, that all products in such expressions involve an \textit{even\/} number of commutators (except the last product in Eq.\ (10) which involves an even number of anticommutators). This fact implies that \textit{all\/} the eigenvectors of the squared Bell operator $\hat{\mathcal{B}}_{\text{M}}^2(n)$ are two-fold degenerate. To see this, let us consider the case of $n$ even, the treatment and conclusions for the case of $n$ odd being the same as for $n$ even. Without loss of generality we may take the axes ${\hat{\textbf {n}}}_j$ and ${\hat{\textbf {n}}}_{j}^{\prime}$ as defining the \mbox{$x$--$y$} plane associated with particle $j$, so that such directions ${\hat{\textbf {n}}}_j$ and ${\hat{\textbf {n}}}_{j}^{\prime}$ are specified by the azimuthal angles $\phi_j$ and $\phi_j^{\prime}$, respectively. Thus, replacing in Eq.\ (10) each commutator $[{\hat{\textbf {n}}}_j,{\hat{\textbf {n}}}_{j}^{\prime}]$ (anticommutator $\{ {\hat{\textbf {n}}}_j,{\hat{\textbf {n}}}_{j}^{\prime} \}$) by its value $2i\sin\theta_j \,{\hat{\textbf {z}}}_j$ ($2\cos\theta_j \hat{I}_j$), we obtain
\begin{multline}
\hat{\mathcal{B}}_{\text{M}}^{2}(\text{$n$ even})=2^{n-1} \Bigl( \hat{I}
+ \sum_{m_1 < m_2} \sin\theta_{m_1}\sin\theta_{m_2}\, {\hat{\textbf {z}}}_{m_1}
{\hat{\textbf {z}}}_{m_2}  \\
+ \sum_{m_1 < m_2 < m_3 < m_4} \sin\theta_{m_1}\sin\theta_{m_2} \sin\theta_{m_3}\sin\theta_{m_4}\, {\hat{\textbf {z}}}_{m_1}
{\hat{\textbf {z}}}_{m_2}{\hat{\textbf {z}}}_{m_3}
{\hat{\textbf {z}}}_{m_4} \\
+\ldots + \sum_{m_1 < m_2 < \ldots < m_{2k}} \sin\theta_{m_1}\sin\theta_{m_2}\ldots \sin\theta_{m_{2k}}\, {\hat{\textbf {z}}}_{m_1}
{\hat{\textbf {z}}}_{m_2}\ldots {\hat{\textbf {z}}}_{m_{2k}} \\
+\ldots + \sum_{m_1 < m_2 < \ldots < m_{n-2}} \sin\theta_{m_1}\sin\theta_{m_2}\ldots \sin\theta_{m_{n-2}}\, {\hat{\textbf {z}}}_{m_1}
{\hat{\textbf {z}}}_{m_2}\ldots {\hat{\textbf {z}}}_{m_{n-2}} \\
+ \sin\theta_1\sin\theta_2\ldots \sin\theta_n\, {\hat{\textbf {z}}}_1
{\hat{\textbf {z}}}_2\ldots {\hat{\textbf {z}}}_n  -(-1)^{n/2}\cos\theta_1\cos\theta_2\ldots \cos\theta_n \hat{I} \Bigr) ,
\end{multline}
where $\theta_j=\phi_j^{\prime}-\phi_j$ is the angle included between ${\hat{\textbf {n}}}_j$ and ${\hat{\textbf {n}}}_{j}^{\prime}$, and ${\hat{\textbf {z}}}_j$ is the spin operator for particle $j$ along its own $z$-axis. From Eq.\ (18), it is apparent that every one of the $2^n$ vectors $|z_1,z_2,\ldots,z_n \rangle$ is an eigenvector of $\hat{\mathcal{B}}_{\text{M}}^{2}(\text{$n$ even})$, where $|z_j \rangle$ designates the eigenvector of ${\hat{\textbf {z}}}_j$ with eigenvalue $z_j =+1 \text{ or }-1$. Also, from Eq.\ (18), it is clear that if $|z_1,z_2,\ldots,z_n \rangle$ is an eigenvector of $\hat{\mathcal{B}}_{\text{M}}^{2}(\text{$n$ even})$ with associated eigenvalue $\mu$, then the same holds true for the eigenvector $|-z_1,-z_2,\ldots,-z_n \rangle$. For the special case where $\theta_j =\pi/2$ for each $j=1,2,\ldots,n$, Eq.\ (18) becomes
\begin{multline}
\hat{\mathcal{B}}_{\text{M}}^{2}(\text{$n$ even})=2^{n-1} \Bigl( \hat{I}
+ \sum_{m_1 < m_2} {\hat{\textbf {z}}}_{m_1}
{\hat{\textbf {z}}}_{m_2}  
+ \sum_{m_1 < m_2 < m_3 < m_4} {\hat{\textbf {z}}}_{m_1}
{\hat{\textbf {z}}}_{m_2}{\hat{\textbf {z}}}_{m_3}
{\hat{\textbf {z}}}_{m_4} \\
+\ldots + \sum_{m_1 < m_2 < \ldots < m_{2k}} {\hat{\textbf {z}}}_{m_1}
{\hat{\textbf {z}}}_{m_2}\ldots {\hat{\textbf {z}}}_{m_{2k}} \\
+\ldots + \sum_{m_1 < m_2 < \ldots < m_{n-2}} {\hat{\textbf {z}}}_{m_1}
{\hat{\textbf {z}}}_{m_2}\ldots {\hat{\textbf {z}}}_{m_{n-2}} 
+ \, {\hat{\textbf {z}}}_1 {\hat{\textbf {z}}}_2\ldots {\hat{\textbf {z}}}_n  \Bigr) .
\end{multline}
Since the total number of terms in (19) is $\sum_{k=0}^{n/2} \binom{n}{2k} =2^{n-1}$, we can immediately conclude that, for the considered case in which $\theta_j =\pi/2$, $j=1,2,\ldots,n$, both \mbox{$|\uparrow \uparrow \cdots \uparrow \rangle$} and \mbox{$|\downarrow \downarrow \cdots \downarrow \rangle$} are eigenvectors of $\hat{\mathcal{B}}_{\text{M}}^{2}(\text{$n$ even})$ with eigenvalue $2^{2(n-1)}$.\footnote{%
It will be noted that each single-particle operator ${\hat{\textbf {z}}}_j$ appears $2^{n-1}/2$ times in Eq.\ (19). As can readily be seen, this very fact makes the eigenvalues corresponding to the $2n$ eigenvectors, $|\downarrow\uparrow\cdots\uparrow\rangle$, $|\uparrow\downarrow\uparrow\cdots\uparrow\rangle, \ldots , $ \mbox{$|\uparrow\cdots\uparrow\downarrow\rangle$}, $|\uparrow\downarrow\cdots\downarrow\rangle$, $|\downarrow\uparrow\downarrow\cdots\downarrow\rangle, \ldots , \text{ and }|\downarrow\cdots\downarrow\uparrow\rangle$ to be zero. Indeed, it can be shown that all $2^n -2$ eigenvalues of the operator in Eq.\ (19) corresponding to the eigenvectors $|z_1,z_2,\ldots,z_n \rangle$ (with $|z_1,z_2,\ldots,z_n \rangle \neq |\uparrow \uparrow \cdots \uparrow \rangle \text{ or }|\downarrow \downarrow \cdots \downarrow \rangle$) are equal to zero.} Obviously, the latter is the maximum eigenvalue of $\hat{\mathcal{B}}_{\text{M}}^{2}(\text{$n$ even})$. (Of course, as noted earlier, the same conclusions hold for the operator $\hat{\mathcal{B}}_{\text{M}}^{2}(\text{$n$ odd})$.) It turns out that the eigenvector $|\Phi^{\pm}\rangle$ of $\hat{\mathcal{B}}_{\text{M}}(\text{$n$ even})$ (or $\hat{\mathcal{B}}_{\text{M}}(\text{$n$ odd})$) with maximum eigenvalue $\lambda_{\text{max}}=\pm 2^{n-1}$ does consist of an equally weighted superposition of the two eigenvectors of $\hat{\mathcal{B}}_{\text{M}}^{2}(\text{$n$ even})$ (or $\hat{\mathcal{B}}_{\text{M}}^{2}(\text{$n$ odd})$) with eigenvalue $\mu_{\text{max}}=2^{2(n-1)}$,
\begin{equation}
|\Phi^{\pm}\rangle = (1/\sqrt{2}) \, ( |\uparrow \uparrow \cdots \uparrow \rangle
\pm e^{i\phi} |\downarrow \downarrow \cdots \downarrow \rangle ) \, ,
\end{equation}
where the phase $\phi$ is given by $\phi = \phi_1 +\phi_2 +\ldots +\phi_n + \pi/2$. (Recall that $\phi_j^{\prime}=\phi_j +\pi/2$ if the Bell inequalities (6) are to be maximally violated by the state $|\Phi^{\pm}\rangle$.) For the Bell operator (7) discussed by Mermin [9] we have $\phi_1 =\phi_2=\ldots=\phi_n =0$, and then $|\Phi^{\pm} \rangle =(1/\sqrt{2}) \left(|\uparrow \uparrow \cdots \uparrow \rangle \pm i |\downarrow \downarrow \cdots \downarrow \rangle \right)$ is the (nondegenerate) eigenvector of operator (7) with eigenvalue $\pm 2^{n-1}$.

In conclusion, by expressing the square of the Mermin's Bell operator in terms of the $n$ single-particle commutators $[{\hat{\textbf {n}}}_j,{\hat{\textbf {n}}}_{j}^{\prime}]$, we have made explicit the relationship between the operators' noncommutativity and the quantum-mechanical violation of the Bell inequality for the general case in which $n$ spin-$\tfrac{1}{2}$ particles are considered. We have seen that in order for the quantum-mechanical predictions to maximally violate Mermin's inequality, it is necessary that all $n$ anticommutators vanish. Furthermore, we have shown that the maximal violation of Mermin's inequality predicted by quantum mechanics decreases exponentially with the number of vanishing commutators. Last, but not least, it is the case that the diagonalisation of the operator $\hat{\mathcal{B}}_{\text{M}}^{2}(n)$ (and hence the diagonalisation of $\hat{\mathcal{B}}_{\text{M}}(n)$ itself) can readily be performed when such Bell operator squared is expressed in terms of the commutators.

\newpage

\center

\end{document}